\documentclass[12pt]{article}
\usepackage{graphicx}

\usepackage{tikz}  
\usepackage{amsmath} 
\usepackage{slashed}
\usepackage{simplewick}
\usepackage{subfigure}
\usepackage{slashed}
\usepackage{simplewick}
\usepackage{fancybox}
\usepackage{cancel}
\usepackage{graphicx}
 \usepackage{setspace} 
    \usepackage{tabu}
   \usepackage{amsmath}
    \usepackage{tikz}
    \usepackage{verbatim}
\usetikzlibrary{arrows,shapes,backgrounds}
\usepackage{mathtools}
\usetikzlibrary{calc}
\usetikzlibrary{positioning}

\pgfmathtruncatemacro\distance{1}

\usepackage{subfigure}
\usepackage{slashed}
\usepackage{simplewick}

\newcommand{\be}{\begin{equation}}
\newcommand{\ee}{\end{equation}}
\newcommand{\bea}{\begin{eqnarray}}
\newcommand{\eea}{\end{eqnarray}}

\newcommand{\lp}{\left(}
\newcommand{\rp}{\right)}

\newcommand{\nn}{\nonumber}
\newcommand{\figref}[1]{Fig.~\ref{fig#1}}
\newcommand{\tableref}[1]{Table~\ref{tab#1}}
\newcommand{\equref}[1]{Eq.~(\ref{#1})}
\newcommand{\ct}[1]{~\cite{#1}}

\newcommand{\tx}[1]{\mathrm{#1}}

\def\gev{\, {\rm GeV}}

\newcommand{\gsim}{\lower.7ex\hbox{$\;\stackrel{\textstyle>}{\sim}\;$}}
\newcommand{\lsim}{\lower.7ex\hbox{$\;\stackrel{\textstyle<}{\sim}\;$}}

\newcommand{\pb}{\rm pb}

\newcommand{\g}{ \gamma}

\newcommand{\ig}[2]{ \raisebox{-0.5\height}{ \includegraphics[width=\textwidth,height=#1\textheight,keepaspectratio]{#2}}  }

\newcommand{\ud}{
	{\renewcommand{\arraystretch}{0.6}
	\begin{tabular}{l} 
	\scalebox{0.8}{$  2 m_\mu \left| {d_\mu \over e} \right| <  1.6 \times 10^{-6} $}     \\ 
	 \scalebox{0.618}{BNL E821 2009} \\
	 \end{tabular}}
	 }

\newcommand{\ua}{	 
	 {\renewcommand{\arraystretch}{0.6}
	\begin{tabular}{l} 
	\scalebox{0.8}{$ \Delta a_{\mu} =  2.87\times 10^{-9}$}   \\ 
	 \scalebox{0.618} { exp. BNL E821 2006} \\
	 \scalebox{0.618} { th. PDG 2013} \\
	 \end{tabular}}
	 }
	 
\newcommand{\ea}{	 
	  {\renewcommand{\arraystretch}{0.6}
	\begin{tabular}{l} 
	\scalebox{0.8}{$\Delta a_e =  -0.37  \times 10^{-12}$ }   \\ 
	 \scalebox{0.618} { exp. DELPHI 2004} \\
	 \end{tabular}}
	 }
	 
\newcommand{\ed}{	 	 
	 {\renewcommand{\arraystretch}{0.6}
	\begin{tabular}{l} 
	\scalebox{0.8} {$ 2 m_e \left| {d_{\text{e}} \over e} \right| < 5.5 \times 10^{-17}$ }   \\ 
	 \scalebox{0.618} {Hudson et al 2011} \\
	 \end{tabular}}
	 }

\usepackage{enumerate}
\usepackage{amssymb,amsmath}

\newcommand\pubnumber{DPF2013-194}
\newcommand\pubdate{\today}

\def\napoli{Department of Physics and Astronomy\\
University of Hawai'i, Honolulu, HI 96822, USA}

\def\Title#1{\begin{center} {\Large #1 } \end{center}}
\def\Author#1{\begin{center}{ \sc #1} \end{center}}
\def\Address#1{\begin{center}{ \it #1} \end{center}}

\newcommand\pubblock{\rightline{\begin{tabular}{l} \pubnumber\\
         \pubdate  \end{tabular}}}
\newenvironment{Abstract}{\begin{quotation}  }{\end{quotation}}
\newenvironment{Presented}{\begin{quotation} \begin{center} 
             PRESENTED AT\end{center}\bigskip 
      \begin{center}\begin{large}}{\end{large}\end{center} \end{quotation}}
\def\Acknowledgments{\bigskip  \bigskip \begin{center} \begin{large}
             \bf ACKNOWLEDGMENTS \end{large}\end{center}}

\def\beq{\begin{equation}}
\def\eeq#1{\label{#1}\end{equation}}
\def\eeqn{\end{equation}}

\def\beqa{\begin{eqnarray}}
\def\eeqa#1{\label{#1}\end{eqnarray}}
\def\eeqan{\end{eqnarray}}

\let\bar=\overbar

\def\Dslash{\not{\hbox{\kern-4pt $D$}}}
\def\dslash{\not{\hbox{\kern-2pt $\del$}}}

\def\ee{e^+e^-}

\def\msb{{\bar{\ssstyle M \kern -1pt S}}}

\begin{document}
\begin{titlepage}
\pubblock

\vfill
\Title{Dipole Moment Bounds on Scalar Dark Matter Annihilation}
\vfill
\Author{ Keita Fukushima}
\Address{\napoli}
\vfill
\begin{Abstract}
We consider a scalar dark matter annihilations to light leptons mediated by charged exotic fermions.
The interactions of this model also adds a correction to dipole moments of light leptons.
In the simplified model, these processes will depend upon the same coupling constants.
The tight experimental bounds on the dipole moments of light leptons will constrain the coupling constants.
Consequently, this bound will then limit the annihilations.
We will produce this dipole moment bounds on the annihilation. 
From this analysis, we report that the bound on annihilation to the electrons is $4.0\times10^{-7}\pb$ (g-2) + $8.8\times 10^{-15}\pb$ (EDM) and the muons is $5.6\times 10^{-4}\pb$ (g-2) + $180\pb$ (EDM), in the limit where the mediator is much heavier than dark matter. 
The parentheses indicate the dipole moment used to obtain the values. 
We note that only the annihilation to muons through a CP-violating coupling is not excluded from indirect detection experiments.
\end{Abstract}
\vfill
\begin{Presented}
DPF 2013\\
The Meeting of the American Physical Society\\
Division of Particles and Fields\\
Santa Cruz, California, August 13--17, 2013\\
\end{Presented}
\vfill
\end{titlepage}
\def\thefootnote{\fnsymbol{footnote}}
\setcounter{footnote}{0}
%


\section{Introduction}
 
Various dark matter experiments yield different observables depending on its detection strategies. 
It is interesting to consider how each observables can be related to one another. 
One way of approaching this questions is to consider a simplified scenario.
For instance, one can write few effective Lagrangians with one mediating particle to describe the interactions of dark matter and the Standard Model.
This formalism is called simplified models~\cite{SimplifiedModels}.
The advantage of simplified models is that the interactions involve only a few quantities such as the mass, the spin, the coupling constant, and the exchange channel ($s$,$t$, or $u$).
Having such a few arbitrary parameters clarifies the analytical relationships of the different observables.
Thus it allows one to capture the essence of physics behind varieties of new physics models.
It is this class of simplified models that we consider in our paper. 

The observables that we'd like to relate in this paper are the correction to the dipole moments and the dark matter annihilation cross section. 
This is because the dipole moments of light leptons have been extremely well measured by the E821 experiment. 
However, there remains to be a $3.5 \sigma$ discrepancy between theory and experiment of the muon magnetic moment~\cite{MuonMomentAnomaly}.
It is interesting to see how this tight constraint might be related to the dark matter annihilation. 
In this context, we necessarily choose an interaction which contributes to both dipole moment correction and annihilation cross sections. 
To fit our needs, we'll consider a dark matter annihilations to light leptons mediated by charged exotic fermions.
In the simplified model, the magnitudes of the processes will depend upon the same coupling constants. 
Consequently, the tight experimental bounds on the dipole moments can be converted to the annihilation cross section bounds.
We will produce this dipole moment bounds on the annihilation. 

We will find that the effective operator separates into two terms. 
First term will contribute to a $CP$-conserving process bounded by the magnetic moment of the lepton. 
Second term will contribute to a $CP$-violating process bounded by the electric dipole moment. 
Furthermore, in the limit we consider, both the dipole moment correction and the annihilation cross section separates into these two terms. 
This will allow us to obtain two independent bounds on the annihilation processes~\cite{Fukushima:2013efa}. 

The outline of this paper is as follows: In section II, we review the dipole moment bounds in the context of new physics. in section III, we introduce our simplified model. In section IV, we explain the separation of the $CP$-conserving and $CP$-violating terms. Finally in section V, we compute our dipole moment constraints on the annihilation cross section.

\newcommand{\1}{Dipole moments bound new physics}
\newcommand{\2}{A simplified annihilation model}
\newcommand{\3}{A simplified model has $CP$-conserving  and $CP$-violating terms}
\newcommand{\4}{Only the $CP$-violating muon channel is observable}

\section{\1}

The fermion photon vertex is given by  $ i{\cal M}= -i e \bar f \Gamma^\mu f \tilde{A}_{\mu} $, where the loop diagram corrections is given by~\cite{Musolf:1990sa},
\bea
\Gamma^\mu &=& \gamma^\mu F_1(q^2)  +{\imath \sigma^{\mu \nu} q_\nu \over 2m} F_2(q^2)	
+{\imath \sigma^{\mu \nu} q_\nu \gamma^5  \over 2m} F_3(q^2)
+(\gamma^\mu q^2 -2m q^\mu) \gamma^5 F_A(q^2) .
\label{1}
\eea
$F_{1,2,3,A}(q^2)$ are the form factors with a dependence on the momentum of the photon $q$.  
This expression reduces to the ordinary photon vertex in the limit $q \to 0$ and $F_1(0)=1$. 
This is required by gauge invariance. 
The subscript $A$ in $F_A$ stands for the anapole moment which we don't consider here. 
As for the last two remaining terms, it is well known that $F_2(0)$ couples to the magnetic field and $F_3(0)$ couples to the electric field. 
Thus these form factors will effectively contribute to the magnetic and electric dipole moments,
\bea
F_2(0)=a ,\tx{\hspace{10pt}}
F_3(0)= 2m_f {d \over |e|}.
\label{2}
\eea
These quantities have been experimentally extremely well measured, and the example is plotted on \figref{1}.
Magnetic moment of the muon\ct{mm} is shown on the left. 
It is plotted with the theoretical SM prediction in gray. 
It is apparent that there remains to be a $3.5\sigma$ deviation\ct{mm} between experiment and theory. 
The electric dipole moment (EDM) of the electron\ct{electron,Moyotl:2011yv} is shown on the right. 
It is converted to the dimensionless quantity $F_3$ so that the overall magnitude may be compared with the magnetic moment bounds. 
The EDM contribution is generated through a $CP$-violating interaction. 
Thus its discovery will point to new physics. 
At present, there is a very tight upper limit. 
\begin{figure}[h]
\centering
\[ 
\tx{ \ig{0.08}{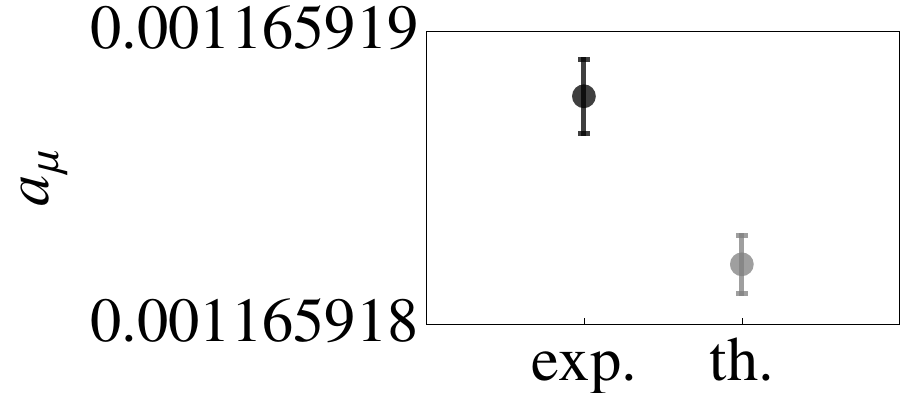} } \Big\}  \tx{\ua}  
\hspace{10pt}
\tx{\ig{0.08}{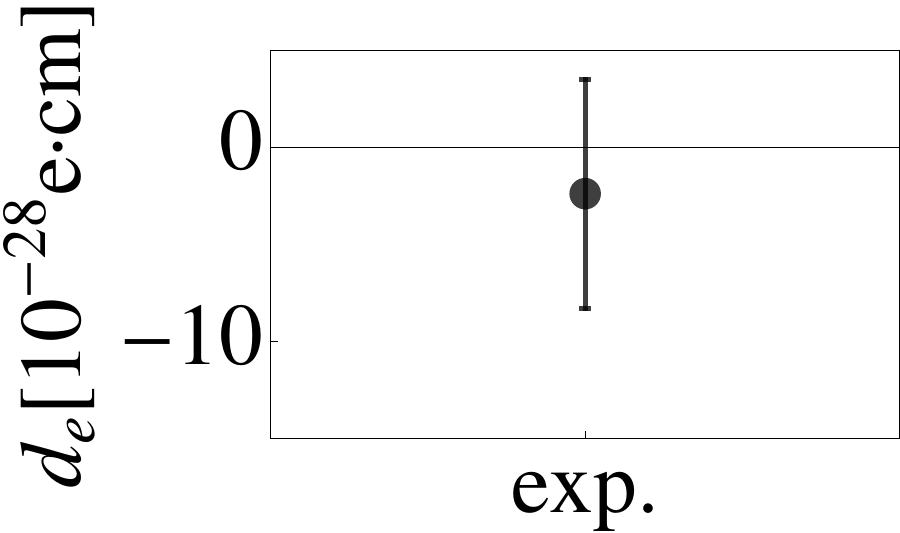} } \big\}\tx{\ed}  \]
\caption{Experimental dipole moment data for the muon g-2 on the left (with SM prediction indicated in gray) and the electron EDM on the right.} 
\label{fig1}
\end{figure}

There are mainly two approaches in describing these discrepancies~\cite{mm}. 
One approach is to improve SM calculation methods by incorporating the experimentally measured strong couplings. 
Another approach is adding new particles running in the loop. 
In this paper we take this latter approach. 
We take the difference of experimental and theoretical value to be the contribution from new physics  as shown in \figref{2}. 
Then the data will bound the Yukawa coupling, which will then bound the annihilation cross section for indirect detections. 
Still, there remains a possibility of a large cancellation between two new physics. 
For example, two new physics contributions may finely cancel with each other to generate the measured data. 
However, we'll assume these scenarios as disfavored and will not be considered. 
This is a presentation that I gave at DPF2013 based on our original paper~\cite{Fukushima:2013efa} . 

\begin{figure}[h]
\centering
\[ 
\Delta a  = \underbrace{{g-2 \over 2} }_{a_{\tx{exp.}}}
-
\underbrace{ 
\ig{0.14}{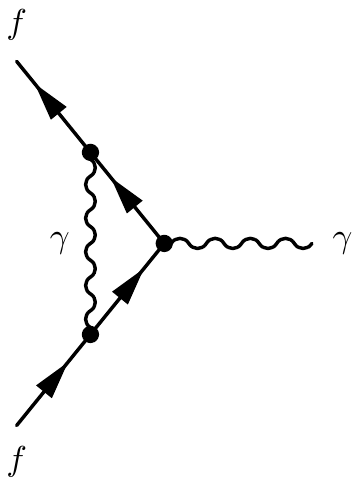}
\ig{0.14}{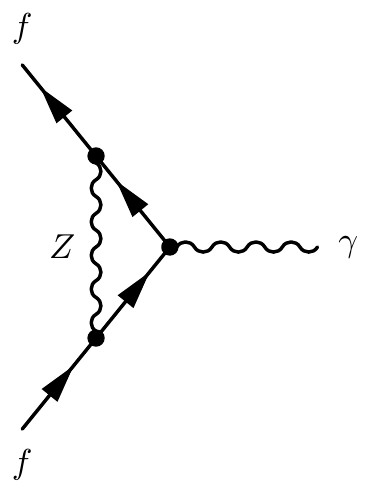}
\ig{0.14}{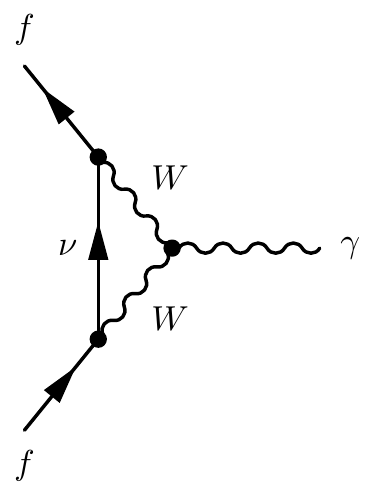}
\ig{0.14}{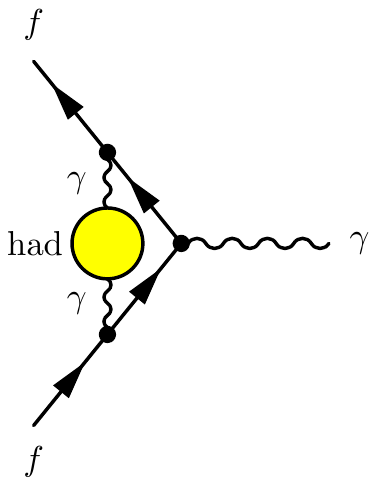}
}_{a_{\tx{SM}}}
= \underbrace{ \ig{0.14}{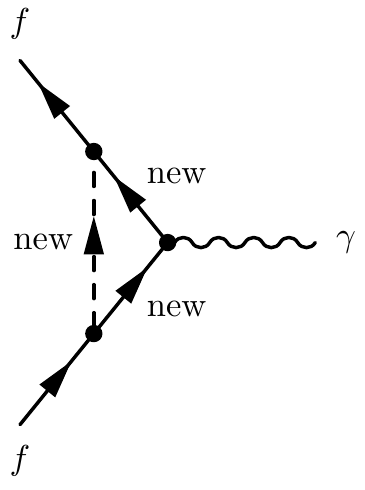}}_{a_{\tx{new}}} 
\]
\caption[]{Muon g-2 anomaly of experiment (1st term) and theory (2nd term). The difference is set equal to the contribution from new physics on the right hand side. }
\label{fig2}
\end{figure}

\section{\2}

Consider a simplified scalar dark matter annihilation model. The essential assumptions are the following:
 \begin{itemize}
      \item The Dark Matter does not decay. It is stabilized under an unbroken symmetry under which it is the lightest particle. Conservation of charge under this symmetry requires it to have a tree level coupling to a heavier mediator.
      \item The dark matter couples to light leptons because leptons which have the tightest experimental dipole moment bounds. Also assuming only leptonic interactions allows one to decouple from direct detection and collider bounds which place constraints on couplings to quarks. 
      \item The mediator is necessarily EM charged so that there are contributions to the dipole moment of the SM leptons to which it couples. S-channel annihilation is possible only if the mediator is neutral. Thus s-channel annihilation will not be constrained.  
    \end{itemize}    
    From these assumptions, the constructed renormalizable Lagrangian becomes  
    \bea 
    {\cal L}_{\tx{int}} = X^* \bar f' \lp \lambda_L P_L+  \lambda_R P_R \rp f  + \tx{h.c.}.
    \label{3}
    \eea 
Examples of models which can realize this interactions include leptophilic dark matter~\cite{LeptophilicDarkMatter},
and this model can be realized in scenarios of WIMPless dark matter~\cite{WIMPless}.

Dark matter $X$ is a exotic complex or real scalar.
The collider searches place tight contraints on couplings to quarks.
In some models, these bounds are avoided by assuming the particles mediating annihilations to quarks to be very heavy. 
If our particles mediating annihilations to leptons are not as heavy, scalar annihilation may still be sizable.  
Also, scalar annihilations does not suffer from chirality/p-wave suppression which arises in the case of Majorana fermions. 
Hence, it is important to consider scalar annihilations.

The mediator $f'$ is a exotic fermion. 
If $X$ is an $SU(2)_L \times U(1)_Y$ singlet, then $f'$ must be chiral under  $SU(2)_L \times U(1)_Y$.
This is required by gauge-invariance.
Hence $f'$ will behave as an exotic lepton which get mass via couplings to higgs. 
So the mass can not be arbitrarily heavy. 
If one further assumes the cancellation of the hypercharge mixed anomaly, it would require the existence of a 4th generation quark.
This case is much more tightly constrained~\cite{4thGenQuarks}.
However, there are other ways to cancel the anomaly such as in presence of another mirror lepton.
We assume here that hypercharge mixed anomaly is cancelled. 

Instead, it is also possible that $X$ is vector-like and non-chiral.
In this case the mass will not be constrained. 
Then, gauge invariance will require that $X$ be a linear combination of field with different charges under $SU(2)_L \times U(1)_Y$ (though electrically neutral).
In this case the the $\lambda_{L,R}$ coefficients will include the relevant mixing angles. 
To summarize, it is possible that $m_{f'}$ is constrained in specific models. 
However for our purposes, we'll assume $m_{f'}$ as unconstrained. 

$\lambda_L$ and $\lambda_R$ is the left- and right-handed fermion Yukawa coupling respectively. 
The overall phase of the $\lambda_L$ and $\lambda_R$ can be absorbed by phase rotations and field redefinitions. 
However, a relative phase between the $\lambda_L$ and $\lambda_R$ may not be absorbed by field redefinition. 
It is this relative phase between the $\lambda_L$ and $\lambda_R$, which generates the $CP$-violating terms which are proportional to $Im( \lambda_L \lambda_R^* )$

\section{\3}

The diagrams constructed from the Lagrangian \equref{3} is shown in \figref{3}.
The main point is that both diagrams share the same two ${\cal L}_{\tx{int}}$ vertexes connected by the mediator propagator. 
\begin{figure}[h]
\centering
\ig{0.21}{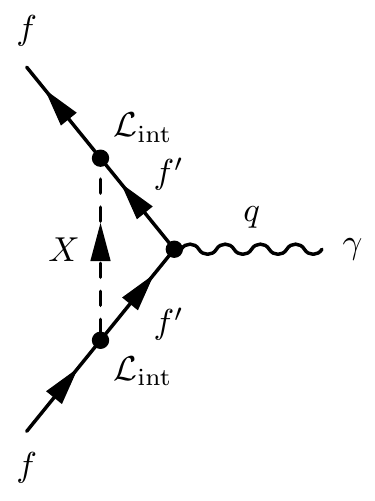} \ig{0.13}{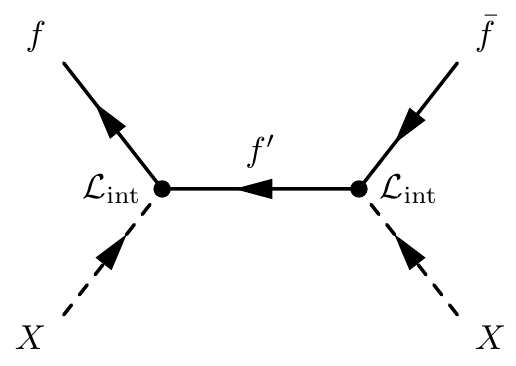}
\caption{The Lagrangian generates a Feynman diagrams of the photon vertex correction (left) and the annihilation (right). Note that both diagrams share this same two vertexes labeled here as ${\cal L}_{\tx{int}}$, connected by the mediator propagator.}
\label{fig3}
\end{figure}
Consequently, the matrix element of the two diagrams will share the same effective operator. 
To achieve this, one can square the ${\cal L}_{\tx{int}}$ and contract the mediator particles. 
The dimension 5 effective operators obtained is,
\bea
 {\cal O} \sim {Re (\lambda_L \lambda_R^*) \over m_{f'} } 
\overbrace{
(X^* X )
}^{\mbox{$CP$ even}}  
\underbrace{
( \bar f f )
}_{\mbox{$CP$ even}} 
+
  { Im (\lambda_L \lambda_R^*) \over m_{f'} } 
\overbrace{
(X^* X)
}^{\mbox{$CP$ even}}
\underbrace{ 
(- i \bar f  \gamma^5 f ) 
}_{\mbox{$CP$ odd}} .
\label{4}
\eea
Here, the limit $m_f \ll m_X \ll m_{f'}$ and $| \lambda_{L,R}| \not \approx 0$ is taken for simplicity.
The overall $CP$ property is labeled with braces. 
The first term is the $CP$-conserving term and the second term is the $CP$-violating term. 
One can observe that the $CP$-conserving term is related to $Re (\lambda_L \lambda_R^*)$ and the $CP$-violating term is related to $Im (\lambda_L \lambda_R^*)$. 

Incidentally, the computed diagrams also separates into the $CP$-conserving and $CP$-violating terms.
Explicitly to lowest order in one-loop correction, the calculated diagrams are~\cite{Cheung:2009fc}, 
\bea
  -i e \bar f \Gamma^\mu f \tilde{A}_{\mu} &\simeq&
  -i e \Big[  
  \bar f  {i \sigma^{\mu \nu} q_\nu \over 2m}  \overbrace{ Re(\lambda_L \lambda_R^*)	{  m_{f} \over m_{f'} } }^{\mbox{$a_\tx{new}$}}  f  
   + 
     \bar f   {\imath \sigma^{\mu \nu} q_\nu \over 2m}   \g^5 \overbrace{   Im(\lambda_L \lambda_R^*)	{  m_{f} \over m_{f'} } }^{\mbox{$2 m {d_\tx{new} \over |e|}$}}  f  
     \Big] \tilde{A}_{\mu},  		 	
     \nn \\
     \sigma v &\simeq&  \Big[           		
			 Re(\lambda_L \lambda_R^*)^2 
			 +   
		Im(\lambda_L \lambda_R^*)^2 
		\Big] {  1\over 4 \pi  m_{f'}^2  }.
		\label{5}
\eea
A careful reader may verify this effective approach with the exact solution in Eq. (\ref{7}-\ref{9}) of Appendix by taking the limit as $m_f \to 0 $. 
These factors are not affected by wave function renormalization of $F_1(q^2)$ to lowest order, which is not taken into account. 
Also, the non-relativistic limit taken here is relevant in the current epoch since the s-wave scalar annihilation is not chirality/p-wave suppressed.

This limit of $m_f \to 0 $ will not apply if either $\lambda_L$ or $\lambda_R$ is sufficiently small.
In this case the leading annihilation cross section terms will scale as $m_f^2$ or $v^4$~\cite{3body} or internal bremsstrahlung processes of three-body final states will become relevant. 
However, these annihilation scenarios will be suppressed and will not be considered here. 
As one can check with \equref{9}, these annihilation terms will not be completely bounded by the two dipole moment bounds, $Re (\lambda_L \lambda_R^*)$ and $Im (\lambda_L \lambda_R^*)$, if either $\lambda_L$ or $\lambda_R \approx 0 $.

On the other hand in the case of large $\lambda_L$ and $\lambda_R$, there is a simple relationship between the the dipole moment correction and the annihilation cross section.
The $CP$-conserving terms are related through the $Re (\lambda_L \lambda_R^*)$ and the $CP$-violating terms are related through $Im (\lambda_L \lambda_R^*)$. 
Indeed it is this bound on the Yukawa couplings, which can be converted to a complete annihilation cross section bounds.

\section{\4}

%
%
%
The experimental bounds~\cite{muon,mm,electron} of the dipole moments are shown in \equref{10}. 
They are ordered in magnitude from left (largest) to right (smallest). 
\bea 
\ud  \hspace{5pt} \ua  \hspace{5pt} \ea  \hspace{5pt} \ed 
\label{10}
\eea
Again, the EDM is converted to the dimensionless $F_3$ form so that it can be compared with the $F_2$ value. 
It is straightforward to see from~\equref{5} that the dipole moment bounds $\Delta a \gtrsim a_{\tx{new}}$ and $2md / |e| \gtrsim 2md_{\tx{new}} / |e|$ in \equref{10} can be converted to annihilation cross section bound as,
\bea
 7.7 \times 10^{11}\pb \left\{ \lp \Delta a_f \rp^2 + \lp 2m_f {d_f  \over |e|} \rp^2 \right\} \left({ \gev \over m_f} \right)^{2} \gtrsim \sigma v  .
\label{6}
\eea
\begin {table}[h]
\caption{Projected dipole moment bounds on scalar dark matter annihilation cross sections to light leptons in the limit, $m_f \ll m_X \ll m_{f'}$ and $| \lambda_{L,R}| \not \approx 0$ }
\begin{center}
\begin{tabular}{c|c|c}
				& CP g-2 bounds	&$\cancel{\tx{CP}}$ EDM bounds							\\
\hline
Electrons $e$	& $4.0 \times 10^{-7} \pb$			& $8.8 \times 10^{-15} \pb $		\\
Muons $\mu$	& $5.6 \times 10^{-4} \pb$			& $180 \pb$ 					\\
Taus $\tau$		& 	non-perturbative		& 		non-perturbative								
\end{tabular}
\end{center}
\label{tab1}
\end{table}
The calculated upper limits are shown in \tableref{1} and \figref{4}. 
These bounds will be multiplied by $\times 4 $ for real scalar annihilation. 
Note that the annihilation cross section scales as the dipole moment squared. 
Thus the tight experimental dipole moment bound is enhanced when converted into the annihilation cross section constraints. 
Also, there is a relative factor of $m_f^2$ between the g-2 and the EDM terms. 
This factor will affect the hierarchy of the magnitudes of the annihilation bounds in some cases.  
	     
The projected bounds of the experiments for $m_X = 10, 100 \gev$ are shown in the $( \langle \sigma_a v \rangle ,m_{f'})$ plane of \figref{4}. 
\begin{figure}[h]
\ig{0.5}{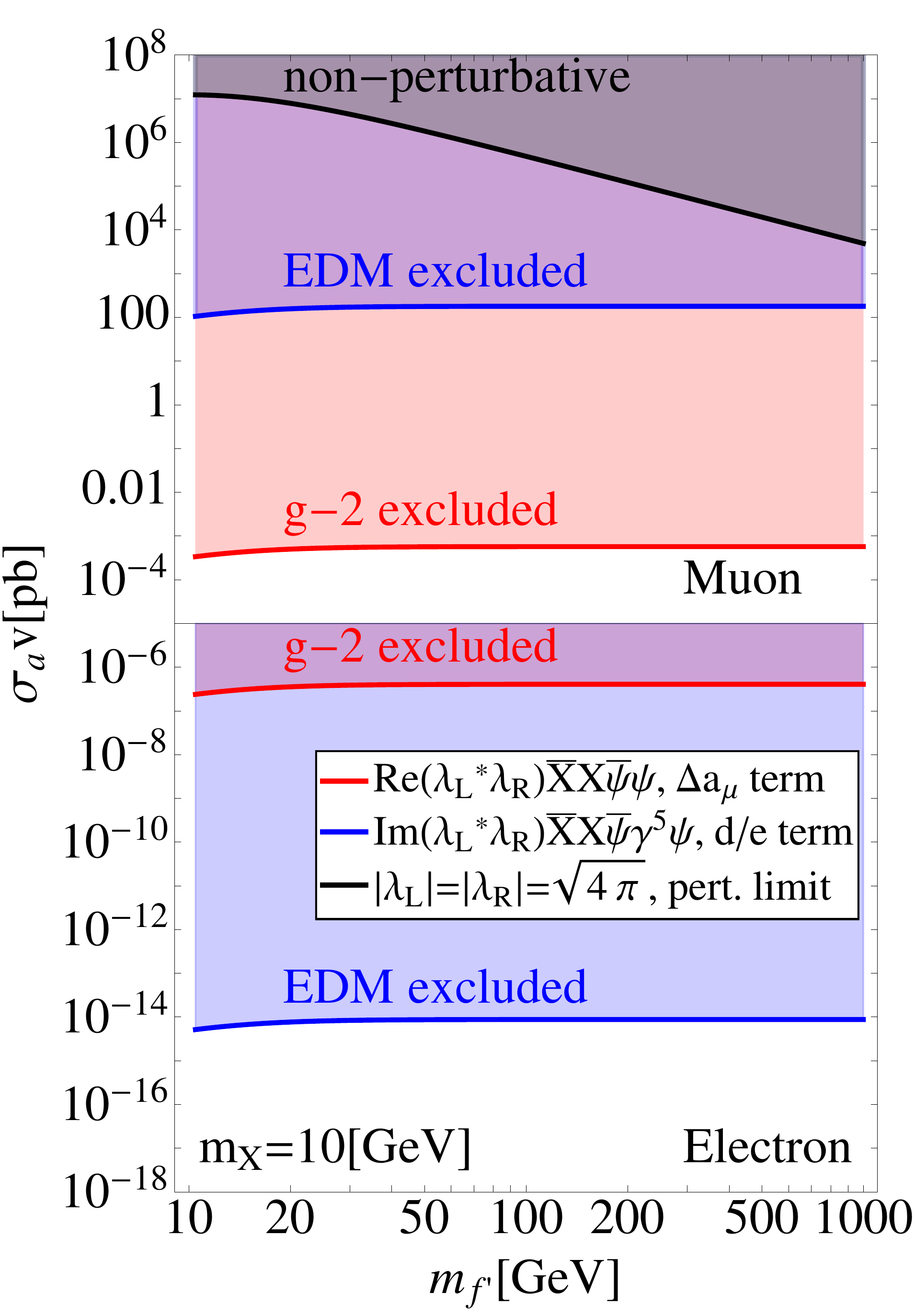} \ig{0.5}{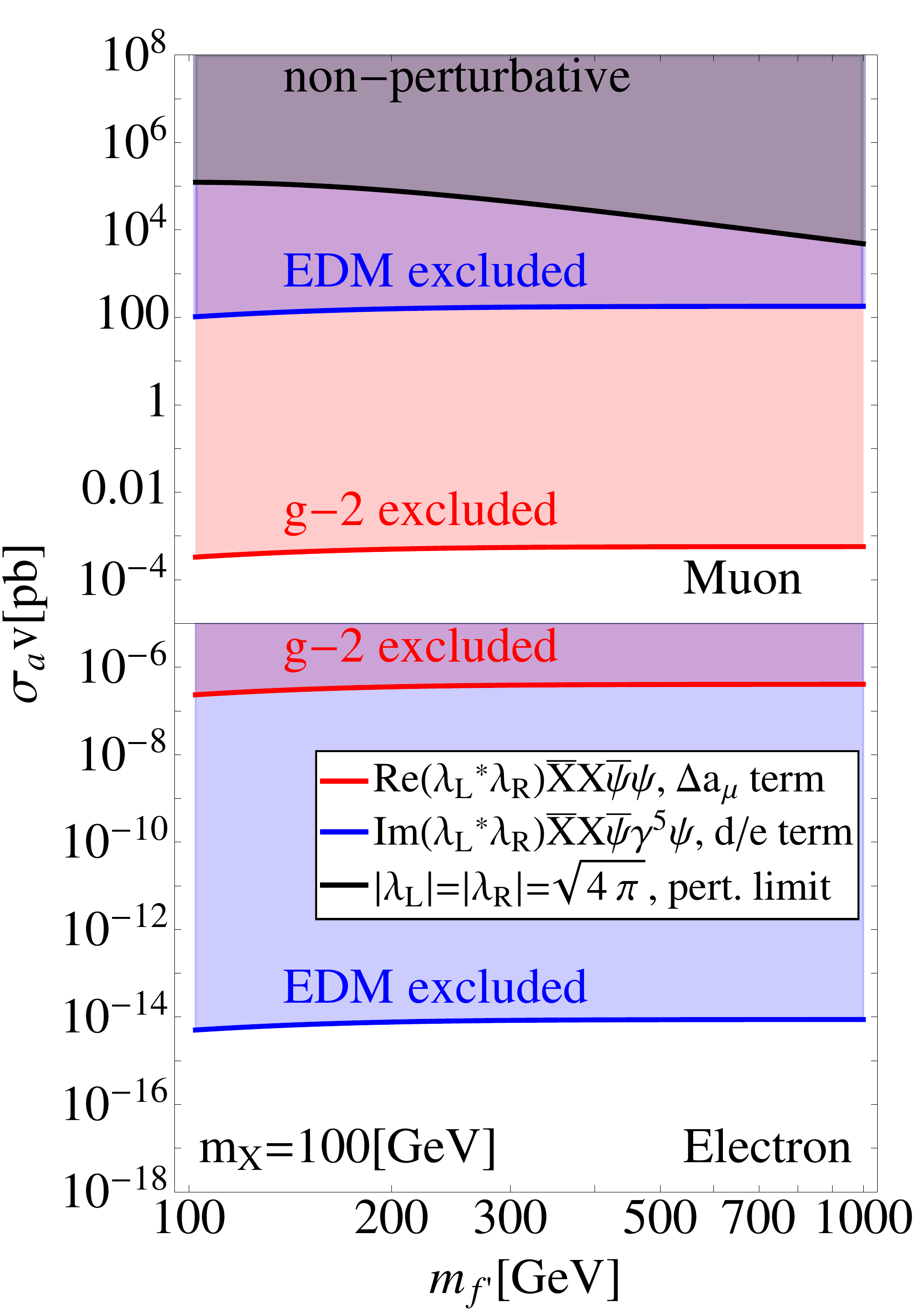}
\caption{The annihilation cross section exclusion plot computed from the dipole moment bounds on the Yukawa couplings for $m_X = 10, 100 \gev$. The top panels show the  $X^* X \to \mu^+ \mu^-$ bounds and the bottom panels show the  $X^* X \to e^+ e^-$ bounds. Red line indicates the $CP$-conserving g-2 bound. Blue line indicates the $CP$-violating EDM bound. The black lines indicates the region where the perturbation variable exceeds {\cal O}(1). The bounds are multiplied by $\times 4 $ for real scalar annihilations.}
\label{fig4}
\end{figure}
The exclusion is indicated by the shaded regions. 
Shown on the top box is the exclusion of annihilation process $X^* X \to \mu^+ \mu^-$ and on the bottom box is the exclusion of annihilation process $X^* X \to e^+ e^-$. 
Shown in red are the g-2 bounds which arises from the $CP$-conserving coupling. 
Shown in blue are the EDM bounds which arises from the $CP$-violating coupling. 
Shown in black are the limit where the perturbation expansion parameter becomes greater than {\cal O}(1). 
The $X^* X \to \tau^+ \tau^-$ bound were not plotted because its bound is above this non-perturbative line. 
The $m_f'$ exclusion from LEP is $\sim100 \gev$ ~\cite{pdg,Carpenter:2011wb}. 
The LHC constraint is tighter by a factor for three if lepton couple's to $SU(2)_L$ and weaker if it does not~\cite{lhcslepton,lhcsleptonrh}.

It is interesting to note that the EDM bound is tighter for the electrons whereas the g-2 bound is tighter for muons. 
Moreover, the current experimental limits on the annihilation cross section to leptons lies around $\sim1\pb$. And the projected limits lies around $10^{-4} \pb$ \cite{Bauer:2013ihz}. 
Therefore, if anything is observed in the range of $10^{-4} \sim 1\pb$ in the future, the there must be a significant contributions from the annihilation to muons that arises from a $CP$-violating coupling in this simplified model. 

It is also interesting to estimate how the scalar annihilation bounds convert to fermion annihilation bounds.
Consider an annihilation of fermions with a t-channel exchange of charged scalar mediator. 
The scaling of the dipole moment bounds coverts from $m_f / m_{f'}$ to $(m_f m_X / m_{f'}^2)  \left\{ (m_f / m_X)\mbox{ or }(\sin \alpha) \right\} $, where $\alpha$ is the scalar mixing angle~\cite{Cheung:2009fc}.
Insertion of the $(m_f / m_X)\mbox{ or }(\sin \alpha)$ factor is necessary when coupling SM leptons of the same helicity states. 
The s-wave annihilation cross section of fermion dark matter goes as $(m_X^2 / m_{f'}^2)$. 
Therefore the resulting dipole moment bounds on fermion annihilation scales roughly the same as the scalar annihilation in the case of maximal mixing (The constraint on the dipole moment would be much weaker if the mixing is negligible~\cite{Buckley:2013sca}).
Similarly, if the fermion is Majorana with negligible mixing,  the s-wave annihilation cross section of fermion dark matter goes as $(m_X^2 / m_{f'}^2)(m_f^2 / m_{X}^2)$.
And the resulting bound is again unchanged from the scalar bound up to an {\cal O }(1) factor.
It is important to note that, however in both cases with the unchanged annihilation bounds, the Yukawa coupling bound is loosened by a factor of $m_{f'} / m_{X}$ or $m_{f'} / m_{f}$ for Dirac and Majorana fermion annihlations respectively. 
Consequently, the bound from non-perturbativity is enhanced by a factor of $m_{f'}^2 / m_{X}^2$ and  $m_{f'}^2 / m_{f}^2$ respectively, which will be significant in cases of heavy mediators.

\section{Conclusion}

We have investigated a simplified model of scalar dark matter with a contribution to both dipole moment correction and an annihilation cross section. 
We have shown that  there is a straightforward relationship between the two processes.
In particular, the bound separated into two terms. 
The g-2 data bounded the $CP$-conserving, $Re(\lambda_L \lambda_R^*)$ term. 
The EDM data bounded the $CP$-violating, $Im(\lambda_L \lambda_R^*)$ term. 
The magnitudes of both processes depended upon these Yukawa couplings.
Consequently, the dipole moment data placed tight constraints on the annihilation cross sections as shown in \tableref{2}.
\begin {table}[h]
\caption{Summary of the results}
\begin{center}
\begin{tabular}{c|c|c}
				& CP g-2 bounds	&$\cancel{\tx{CP}}$ EDM bounds							\\
\hline
Experimental bound 	& $\Delta a \gtrsim Re(\lambda_L \lambda_R^*)	{  m_{f} \over m_{f'} }  $ & $2m{d \over e} \gtrsim Im(\lambda_L \lambda_R^*)	{  m_{f} \over m_{f'} } $ \\
Annihilation cross section	to...	& $Re(\lambda_L \lambda_R^*)^2  {  1\over 4 \pi  m_{f'}^2  }$		&$Im(\lambda_L \lambda_R^*)^2  {  1\over 4 \pi  m_{f'}^2  }$ \\
Electrons $e$	& $4.0 \times 10^{-7}\pb$			& $8.8 \times 10^{-15} \pb$		\\
Muons $\mu$	& $5.6 \times 10^{-4}\pb$			& $180\pb$				\\
Taus $\tau$		& 	non-perturbative		& 		non-perturbative												
\end{tabular}
\end{center}
\label{tab2}
\end{table}
It is interesting to note that the g-2 bound was tighter for the annihilation to muons $X^*X \rightarrow \mu^+ \mu^-$, whereas the EDM bound was tighter for the annihilation to electrons $X^*X \rightarrow e^+ e^-$. 
Furthermore, we found that all calculable annihilation cross sections will be excluded except the annihilation cross section to muons $X^*X \rightarrow \mu^+ \mu^-$ that arises from a $CP$-violating interaction. 
We have also estimated that in the case of MSSM neutralino, we expect the bounds on annihilation to muons $X^* X \to \mu^+ \mu^-$ to be similar to the $CP$-conserving case, up to {\cal O}(1) factors. 
Therefore the branching fraction for neutralino annihilations to muons in the current epoch will still be quite small.

There were several ways in which these constraints of the simplified model can be avoided. 
First, the simplified models consisted of annihilations mediated through charged exotic fermions. 
Then the t- and u-channel is constrained but s-channel is not constrained here. 
Second, we have taken the limits $m_f \ll m_X, m_{f'}$ and $| \lambda_{L,R}| \not \approx 0$. So if either of the coupling constant is zero, the constraint does not apply. 
Third, if there are large cancellation between new physics then larger values of the annihilation cross sections will be allowed. 
Fourth, we have not considered the p-wave suppressed annihilations.  
Lastly, annihilation to $\bar \tau \tau$ are still unconstrained and $\bar q q$ was not evaluated because the experimental dipole moment constraints are much weaker. 
It will be interesting to come back and evaluate to these cases in the future. 


\Acknowledgments
We are grateful to M.~Buckley, D.~Hooper and X.~Tata for useful discussions.
K.~F~thanks the Meeting of the American Physical Society(APS) Division of Particles and Fields(DPF 2013)
for their support and hospitality for the completion of this presentation.

\appendix
\label{appendix}

\section{Exact expressions of the photon vertex correction and the annihilation cross section}

\bea
 F_2 (0)  = {1 \over (4 \pi)^2 }  \int_0^1 dz    {  - (|\lambda_L|^2   + |\lambda_R|^2 ) {1\over2}z(1-z)^2m_f   + (\lambda_L \lambda_R^*   + \lambda_R \lambda_L^*  )(1-z)^2 m_{f'} \over  (1-z) (m_{f'}^2-zm_{f}^2 )  + z m_X^2}  
\label{7}
\eea

\bea
 F_3 (0)  =  {1 \over (4 \pi)^2 }  \int_0^1 dz     {    (\lambda_L \lambda_R^*   - \lambda_R \lambda_L^* )(1-z)^2 m_{f'} \over  (1-z) (m_{f'}^2-zm_{f}^2 )  + z m_X^2}  
\label{8}
\eea

  \bea
(\sigma | v_A - v_B |)_{CM}
 = &&
 - {  \sqrt{1- {m_{f}^2 \over m_{X}^2}} \over 64 \pi m_X^2 (-m_{f}^2 + m_{f'}^2 + m_{X}^2)^2  }
\nn \\ && \times
\Big\{
(\lambda _L \lambda _L^*+\lambda _R \lambda _R^* )^2   \Big[   m_{f}^2 (m_{f}-m_{X}) (m_{f}+m_{X}) \Big]
        \nn \\ &&
 + (\lambda _L^* \lambda _R+\lambda _L \lambda _R^* ) (\lambda _L \lambda _L^*+\lambda _R \lambda _R^* )  \Big[  2 m_{f} (m_{f}-m_{X}) (m_{f}+m_{X}) m_{f'}  \Big] 
        \nn \\ &&
        \nn \\ &&
 +(\lambda _L \lambda _L^* \lambda _R \lambda _R^* ) \Big[  2 m_{f'}^2 (m_{f}^2 - 2m_{X}^2) \Big]
        \nn \\ &&
 +  (\lambda _L^*{}^2  \lambda _R^2+\lambda _L^2 \lambda _R^*{}^2  ) \Big[   m_{f'}^2 m_{f}^2  \Big] 
  \Big\}
\label{9}
  \eea

\end{document}